\documentclass[twocolumn,showpacs,10pt,aps,amssymb,amsmath,pre]{revtex4}

\usepackage{epsfig}

\newcommand{\be}{\begin{equation}}
\newcommand{\ee}{\end{equation}}
\newcommand{\bea}{\begin{eqnarray}}
\newcommand{\eea}{\end{eqnarray}}

\begin{document}

\title{Synchronized flow and wide moving jams from balanced vehicular traffic}

\author{Florian Siebel and Wolfram Mauser}
\affiliation{
Department of Earth and Environmental Sciences, University of Munich, Luisenstra\ss e 37, D-80333 Munich, Germany
}

\begin{abstract}
Recently we proposed an extension to the traffic
model of Aw, Rascle and Greenberg. The extended traffic
model can be written as a hyperbolic system of balance laws and 
numerically reproduces the reverse $\lambda$ shape of the fundamental 
diagram of traffic flow. In the current work we analyze the steady state 
solutions of the new model and their stability properties. In addition to 
the equilibrium flow curve the trivial steady state solutions form two  
additional branches in the flow-density diagram. We show that the 
characteristic structure excludes parts of these branches resulting in 
the reverse $\lambda$ shape of the flow-density relation. The upper branch is
metastable against the formation of synchronized flow for intermediate
densities and unstable for high densities, whereas the lower branch is
unstable for intermediate densities and metastable for high
densities. Moreover, the model can reproduce the
typical speed of the downstream front of wide moving jams. It further
reproduces a constant outflow from wide moving jams, which is far below the
maximum free flow. Applying the model to simulate traffic flow at a bottleneck
we observe a general pattern with wide moving jams traveling through the
bottleneck. 
\end{abstract}
\date{\today}
\maketitle

%%%%%%%%%%%%%%%%%%%%%%%%%%%%%%%%%%%%%
\section{Introduction}
%%%%%%%%%%%%%%%%%%%%%%%%%%%%%%%%%%%%%
Modeling vehicular traffic flow using methods from continuum fluid dynamics
has a long history~\cite{KuM97,Hel97,HHST01,Hel01,Ker04}, with many
contributions mainly from traffic engineers, physicists and mathematicians. 
The existing models can be subdivided into first and higher order models
according to the highest derivative appearing in the partial differential
equations describing traffic flows~\cite{Joh82}.~\footnote{The
  system studied in this paper is a first order system consisting of
two equations, see Eqs.~(\ref{rhob})-(\ref{vb}).}
First order models, as the classical model of Lighthill, Whitham and 
Richards~\cite{LiW55,Ric56}, 
approximate the higher order models by neglecting diffusion terms 
in the same way the Euler equation approximates the Navier-Stokes
equation. As a consequence, discontinuous data can develop, which have to be
dealt with by a numerical algorithm.
 
In the existing literature on first order systems the role of a source term 
in the partial differential equations describing vehicular
traffic flow on a road section without entries and exits has not been studied 
in depth. Instead, the main focus has been laid on the principal part of the
equations~\cite{AwR00}, i.e. the collection of terms in the partial 
differential equation containing derivatives of order equal to the order of 
the partial differential equation~\cite{DCZ02}, and systems with constant 
relaxation time~\cite{Gre01,Ras02,JiZ203,GKR03,Gre04}. 
In~\cite{SiM205}, we presented
the balanced vehicular traffic model (BVT model), which generalizes the model
of Aw, Rascle and Greenberg~\cite{AwR00,Gre01} by prescribing a more general
source term subsumed under an effective relaxation coefficient. 
Unlike in earlier studies, this effective relaxation coefficient depends on
both traffic density and velocity. As we showed in numerical simulations 
in~\cite{SiM205}, the model can reproduce the observed reverse $\lambda$ shape 
of the fundamental diagram of traffic flow~\cite{KIO83}. 

In the current work we aim at obtaining a better understanding of the 
numerical results of the BVT model. In particular we study the appearance 
of the new branches in the fundamental diagram in the congested regime, which 
finally form the reverse $\lambda$ shape. Our explanation of the reverse
$\lambda$ shape differs from earlier explanations, which explained the reverse
$\lambda$ as a part of a single equilibrium flow branch. In contrast, 
as we will show, the (meta-) stable curve sections of three
steady state branches form the reverse $\lambda$ in the BVT model. Hence, 
the criticism of the one-dimensionality of steady states of fluid-dynamical 
models~\cite{Ker04} does not apply to the BVT model. 

In the BVT model traffic flow is described by the following system of balance 
laws determining the density $\rho = \rho(t,x)$ and velocity $v=v(t,x)$ of 
vehicles
\bea
\label{rhob}
\frac{\partial \rho}{\partial t} + \frac{\partial (\rho v)}{\partial x} & = &
0, \\
\nonumber
\frac{\partial (\rho(v-u(\rho))}{\partial t} + \frac {\partial (\rho v
  (v-u(\rho)))}{\partial x} & = & \\
\beta(\rho,v) {\rho (u(\rho)-v)}. &&
\label{vb}
\eea
As usual, $(t,x)$ denote the time and space variable. $u(\rho)$ denotes 
the {\it equilibrium velocity}, which fulfills 
\bea
\label{ucond1}
u'(\rho) & < & 0 \mbox{~for~} 0 < \rho \le \rho_m, \\
\label{ucond2}
\frac{d^2 (\rho u(\rho))}{d \rho^2} & < & 0 \mbox{~for~} 0 < \rho \le \rho_m.
\eea
The effective relaxation coefficient $\beta(\rho,v)$ fulfills
\bea
\label{beta1}
\beta(\rho,v) < 0 {\rm ~for~} 0 < \rho_1 < \rho < \rho_2 \le \rho_m, v=u(\rho),\\ 
\label{beta2}
\beta(\rho,v) \ge 0 {\rm ~for~} 0 \le \rho \le \rho_1 \ {\rm or} \ \rho_2 \le \rho \le \rho_m,
v=u(\rho), \\
\label{beta3}
\lim_{v \to 0,u_m=u(0)} \beta(\rho,v) \ge 0.  \hspace{1cm}
\eea
Note that in moving observer coordinates, Eqn.~(\ref{vb}) reduces to 
\be
\frac{d}{dt} (v-u(\rho)) = - \beta(\rho,v) (v-u(\rho)),
\ee
i.e. $\beta$ can be interpreted as decay parameter. As an effective parameter,
which takes into account the actual relaxation time but also the reaction
time, the parameter can become negative for intermediate to high 
densities~(\ref{beta1}) (see~\cite{SiM205}). Note that for $\beta < 0$, the 
average velocity $v$ departs from the equilibrium velocity $u(\rho)$, 
i.e. $|v-u(\rho)|$ increases with time. For $\beta > 0$, drivers approach 
the equilibrium velocity $u(\rho)$ with a rate determined by the value of 
$\beta$.
 
In~\cite{SiM205} we used parameter functions and values which
describe traffic flow only qualitatively. In the current work we use the 
equilibrium velocity of~Newell~\cite{New61} 
\be
u(\rho) = u_m\Big(1-\exp\Big(-\frac{\lambda}{u_m}\Big(\frac{1}{\rho} -
\frac{1}{\rho_m}\Big)\Big)\Big)
\ee
with parameter values $u_m = 160 \ {\rm km/h}$, $\lambda = 3600 \ {\rm [1/h/lane]}$, $\rho_m = 160 \ {\rm [1/km/lane]}$ and an effective relaxation coefficient
\be
\label{beta}
\beta(\rho,v)=  \left\{
\begin{array}{ll}
\frac{a_c}{u-v},& \mbox{if~} \tilde{\beta}(\rho,v)(u(\rho)-v) - a_c \ge 0, \\
\frac{d_c}{u-v},& \mbox{if~} \tilde{\beta}(\rho,v)(u(\rho)-v) - d_c \le 0, \\
\tilde{\beta}(\rho,v),& \mbox{else},
\end{array} \right.
\ee
\be
\label{betatilde}
\tilde{\beta}(\rho,v) = \frac{1}{\hat{T} u_m} \Big( |u(\rho) - v + \alpha_1 \Delta
v(\rho)| + \alpha_2 \Delta v(\rho) \Big)
\ee
and
\be
\label{deltav}
\Delta v(\rho) = \tanh \Big( \alpha_3 \frac{\rho}{\rho_m} \Big) \Big( u(\rho) + c
\rho_m \Big(\frac{1}{\rho} - \frac{1}{\rho_m}\Big) \Big),
\ee
with parameters $a_c = 2 \ {\rm m/s^2}$, $d_c = - 5 \ {\rm m/s^2}$, 
$\hat{T} = 0.1 \ {s}$, $\alpha_1 = - 0.2$, $\alpha_2 = - 0.8$, $\alpha_3 = 7$
and $c = -14 \ {\rm km/h}$. The density values, which determine the sign
of $\beta$ according to Eqs.~(\ref{beta1})-(\ref{beta2}), are $\rho_1=19.09 
\ {\rm [1/km/lane]}$
and $\rho_2=\rho_m$. Throughout this work, we model two-lane sections of a
highway without entries and exits with the above parameter set. 
In comparison to~\cite{SiM205} theses parameters describe traffic flow 
more realistically, although we have not used
experimental traffic data to determine them for a specific highway
section. Note that the general conditions~(\ref{beta1})-(\ref{beta3}) are
sufficient to obtain multivalued fundamental diagrams. However, the 
quantitative details depend on the precise choice of $\beta$. As our 
simulations show, the analytically derived properties of the model of Aw, 
Rascle and Greenberg, that the velocity does not become negative and 
collisions do not occur, carry over to our system. For the numerical
simulations of the model equations~(\ref{rhob})-(\ref{vb}) we used a 
high-resolution shock-capturing scheme 
with an approximate Riemann solver. We chose a spatial resolution of 20 m and
dynamically adapted the temporal resolution to half the value obtained from
the Courant condition. The numerical method is described in
detail in~\cite{SiM205}.  

In order to obtain a deeper insight into the structure of the BVT model we
study the smooth steady state solutions in Sec.~\ref{sec:steadystate}. Our
numerical simulations produce more general solutions approximating steady
states solutions, which will be discussed in Sec.~\ref{sec:quasisteadystate}. 
In Sec.~\ref{sec:stability}, we assess the stability properties of the steady
state solutions. With theses results we classify the traffic states of
the BVT model according to the three traffic phases of Kerner~\cite{Ker04} in 
Sec.~\ref{sec:kerner} and apply the BVT model to simulate traffic flow at a
bottleneck in Sec.~\ref{sec:bottleneck}. We conclude the paper in 
Sec.~\ref{sec:discussion}.
%%%%%%%%%%%%%%%%%%%%%%%%%%%%%%%%%%%%%%%%%%%%%%%%%%%%%%%%
\section{Smooth steady state solutions of the BVT model}
\label{sec:steadystate}
%%%%%%%%%%%%%%%%%%%%%%%%%%%%%%%%%%%%%%%%%%%%%%%%%%%%%%%%
For smooth solutions the balance equations describing traffic
flow~(\ref{rhob})-(\ref{vb}) can be rewritten as
\bea
\label{rho}
\frac{\partial \rho}{\partial t} + v \frac{\partial \rho}{\partial x}  +
\rho \frac{\partial v}{\partial x} & = & 0,\\
\label{v}
\frac{\partial v}{\partial t} + (v + \rho u'(\rho)) \frac {\partial
  v}{\partial x}  & = &  \beta(\rho,v) (u(\rho)-v).
\eea
In the following, we study the smooth steady state solutions of the BVT
model. In comparison to the study by Lee, Lee and Kim~\cite{LLK04}, 
our analysis - being performed on a first order system - is considerably simpler. Note, 
however, that due to the possibility of dealing with discontinuous solutions, 
we can in principle have more general solutions in the balanced system
  (\ref{rhob})-(\ref{vb})~(see also~\cite{KuM97} and references therein). 

Let us repeat, that for a steady state solution, there is a coordinate system
$(\tilde{t},z)$ and a constant velocity $w$ 
\bea
x & = & z - w \tilde{t}, \\
t & = & \tilde{t},
\eea 
such that
\bea
\frac{\partial \rho}{\partial \tilde{t}} & = & 0, \\
\frac{\partial v}{\partial \tilde{t}} & = & 0.
\eea
It follows from the continuity equation~(\ref{rho}) that for all steady state
solutions there is a constant $q$ such that
\be
\label{stationary}
\rho v = q + \rho w.
\ee 
Hence steady state solutions are restricted to straight lines in the 
fundamental diagram of traffic flow. Moreover, the minimum and maximum speed
of information propagation in system (\ref{rhob})-(\ref{vb}) limit the
physically admissible steady state solutions, i.e. the velocity $w$, as
\be
\label{characteristiccondition}
\lambda_1 = v + \rho u'(\rho) \le w \le \lambda_2=v.
\ee
Let us assume that $v - w > 0$ (else $q=0$).
Then we can solve Eqn.~(\ref{stationary}) for the density 
\be
\rho = \frac{q}{v-w}
\ee
and substitute it into Eqn.~(\ref{v}), yielding the ordinary differential 
equation for steady state solutions in the BVT model
\be
\label{ode}
\Big(\lambda_{1} - w \Big) \frac{dv}{dz} = \beta(\frac{q}{v-w},v) \Big( u(\frac{q}{v-w}) -v \Big).
\ee
%%%%%%%%%%%%%%%%%%%%%%%%%%%%%%%%%%%%%%%%%%%
\subsection{Trivial steady state solutions}
%%%%%%%%%%%%%%%%%%%%%%%%%%%%%%%%%%%%%%%%%%%
We first study the trivial (constant) solutions of this ODE,
i.e. solutions fulfilling $\frac{dv}{dz}=0$. The solutions are:
\begin{itemize}
\item The equilibrium velocity curve:
\be
v = v^{e} = u(\rho).
\ee
\item The {\it jam line} (compare to the line J of Kerner~\cite{Ker04}): 
\bea
\label{jam}
v = v^j = u(\rho) + (\alpha_1 + \alpha_2) \Delta v(\rho) \\ \nonumber
 {\rm~for~} \rho_1 < \rho < \rho_2.
\eea
\\
Note that $\rho v < \rho u(\rho)$.
\item The {\it high-flow branch:}
\bea 
\label{tip}
v = v^h = u(\rho) + (\alpha_1 - \alpha_2) \Delta v(\rho) \\ \nonumber 
{\rm~for~} \rho_1 < \rho < \rho_2. 
\eea
\\
Note that $\rho v > \rho u(\rho)$ in this case.
\end{itemize}
\begin{figure}[htpb]
\vspace{0.5cm}
\includegraphics[width=0.9 \linewidth]{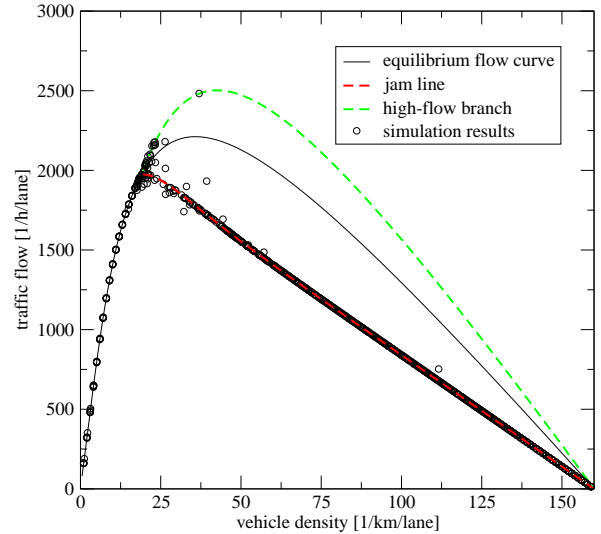}
\vspace{0.3cm}
\caption{(Color online) The trivial steady state solutions (i.e. solutions with 
$\frac{dv}{dz} = 0$) of the BVT model. These solutions are the equilibrium 
solution $v = u(\rho)$ (solid black curve) and the two branches
fulfilling $\beta(\rho,v) = 0$ (dashed curves). Moreover, we present 
the results of simulation runs of perturbed equilibrium data. 
The simulated data points in the fundamental diagram are closely related 
to sections of trivial steady state solutions fulfilling $\beta(\rho,v) = 0$.
\label{fig1}}
\end{figure}
We summarize the trivial steady state solutions in Fig.~\ref{fig1}. We further
show in this figure the results of simulation runs for constant 
initial data in equilibrium $\rho = \rho_0$, $\rho_0 = 1,2,...,159$
[1/km/lane],  
$v = u(\rho_0)$ on a 7 km long stretch of a highway with periodic boundary 
conditions, prescribing a small amplitude perturbation of the density 
$\delta \rho = \sin(\pi x)$ initially located between 2 and 3 km on top. The 
data points were extracted from the simulations at 5 equidistantly 
distributed virtual detectors after an evolution time of 10 h, without
applying a temporal aggregation. As one can 
see from the plot many data points of the numerical solutions are closely 
related to branch sections of trivial steady state solutions, in
particular for the jam line.
%%%%%%%%%%%%%%%%%%%%%%%%%%%%%%%%%%%%%%%%%%%%%%%
\subsection{Non-trivial steady state solutions}
%%%%%%%%%%%%%%%%%%%%%%%%%%%%%%%%%%%%%%%%%%%%%%%
Let us compare the equation of steady states~(\ref{ode}) in analogy to the
dynamics in classical mechanics presented in~\cite{LLK04} to the following
equation
\be
(\lambda_1 - w) \frac{dv}{dz} = - \frac{d}{dv}U(v,w,q),
\ee
with a potential energy $U$. We find that the potential $U(v,...)$ has a 
functional form which is {\it camelback-shaped} for a wide range of constant 
values $w$ and $q$ as in~\cite{LLK04}. However, unlike in~\cite{LLK04}, 
we do not obtain an acceleration term in our first order system. Note that
according to~(\ref{characteristiccondition}) $\lambda_1 -w \le 0$.
 
\begin{figure}[htpb]
\vspace{0.9cm}
\includegraphics[width=0.9\linewidth]{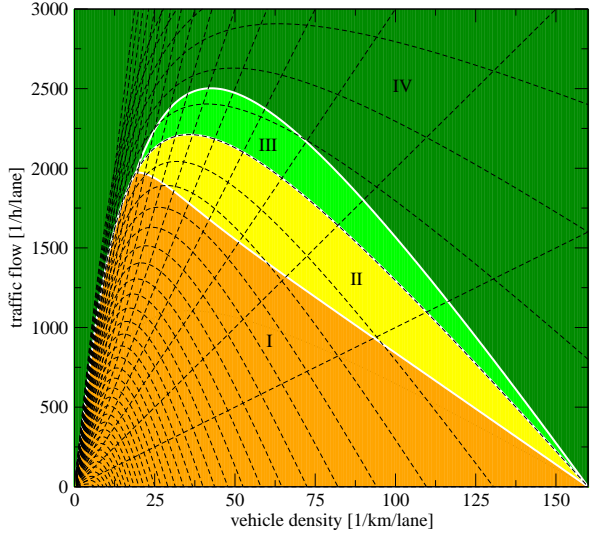}
\vspace{0.2cm}
\caption{(Color online) Characteristic structure and steady state solutions of the BVT
  model. As solid
  lines we plot the trivial steady state solutions bordering the regions I -
  IV. On top (dashed curves) we plot the characteristic curves with 
  slope $\lambda_1$ and $\lambda_2$ respectively. As the speed of steady state 
  solutions is limited by the characteristic speeds $\lambda_1$ and 
  $\lambda_2$, physically admissible steady state solutions lie inside the 
  {\it characteristic cones}
  spanned by these two speeds at every point ($\rho$,$\rho v$) in the 
  flow-density diagram.  
\label{fig2}}
\end{figure}
In the following we restrict the discussion to the physically admissible 
smooth solutions (in particular we do not consider solutions with infinite
gradient). In the limit $z \to \pm \infty$ all maximally extended steady 
state solutions approach one of the following curve sections:
\begin{description}
\item[A:] free equilibrium flow: $v = u(\rho)$ and $\rho \le \rho_1$,
\item[B:] unstable equilibrium flow: $v = u(\rho)$ and $\rho_1 < \rho \le \rho_2$,
\item[C:] jam line: see Eqn.~(\ref{jam}),
\item[D:] high-flow branch: see Eqn.~(\ref{tip}),
\end{description}
We classify the non-trivial maximally extended steady solutions according to
the behavior in the limit $z \to \pm \infty$, using the letters of the
corresponding branches. Due to the
characteristic structure of the BVT model (see Fig.~\ref{fig2}), only five
solution classes can appear. These are the classes AD, BC, BD, CC and
DD~\footnote{For the parameters used in~\cite{SiM205}, for which
  $\rho_2<\rho_m$, we would have an additional
  branch section (E) of stopped equilibrium flow $v = u(\rho)$ for $\rho_2 <
  \rho \le \rho_m$ and a sixth solution class DE.}.
\begin{figure}[htpb]
\includegraphics[width=\linewidth]{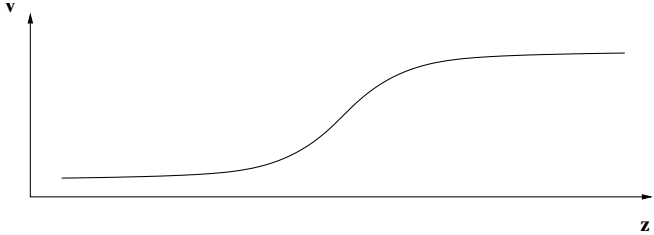}
\caption{Sketch of the non-trivial steady state solutions. The steady state
  solutions fulfill $\frac{dv}{dz} > 0$ (or $\frac{dv}{dz} < 0$
respectively) and $\lim_{z \to \pm \infty} \frac{dv}{dz} = 0$.
\label{fig3}}
\end{figure}
Schematically, all these steady state solutions have the form indicated in
Fig.~\ref{fig3}, i.e. solutions lying in the regions II and IV of 
Fig.~\ref{fig2} fulfill the
condition $\frac{dv}{dz} > 0$, solutions in regions I and III obey 
$\frac{dv}{dz} < 0$ respectively and $\lim_{z \to \pm \infty} 
\frac{dv}{dz} = 0$.
Comparing our classification to the classification of Lee, Lee,
Kim~\cite{LLK04} for cases, where the parameters $q$
and $w$ lead to a camelback-shaped profile of the potential $U$, the solutions
correspond to minimum-saddle solutions (without oscillations). The possible
appearance of the different solution classes in the flow-density diagram is
shown in Fig.~\ref{fig4}.
\begin{figure}[htpb]
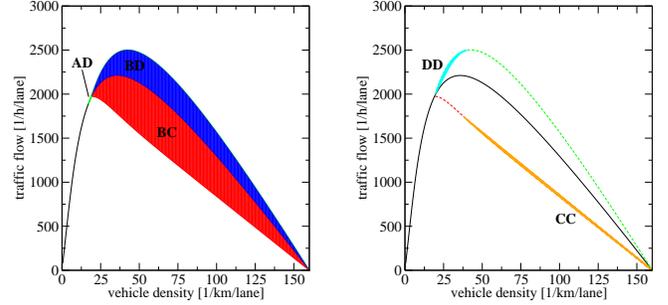

\vspace{0.8cm}
\noindent
\begin{minipage}[htpb]{0.46\linewidth}
\epsfig{figure=fig4a.eps,width=\linewidth}
\end{minipage}\hfill
\begin{minipage}[htpb]{0.46\linewidth}
\epsfig{figure=fig4b.eps,width=\linewidth}
\end{minipage}
\vspace{0.3cm}\\
\caption{(Color online) Left Panel: Regions covered by the smooth maximally extended 
non-trivial steady state
  solutions linking different branches, which are consistent with the
  characteristic structure of the BVT model. The solutions link the free
  equilibrium flow solution with the high-flow branch (class AD), 
  the unstable equilibrium solution with the jam line (class BC) and the 
  unstable equilibrium solution with the high-flow branch (class
  BD).\\
  Right Panel: Regions 
  covered by the smooth maximally extended non-trivial steady state
  solutions linking identical branches, which are consistent with the
  characteristic structure of the BVT model. For class CC the non-trivial
  steady states link data points lying on the jam line, for the class DD 
  they link data points of the high-flow branch.}
\label{fig4}
\end{figure}
%%%%%%%%%%%%%%%%%%%%%%%%%%%%%%%%%%%%%%%%%%%%%%%%%%%%%%%%%%
\section{Quasi steady state solutions}
\label{sec:quasisteadystate}
%%%%%%%%%%%%%%%%%%%%%%%%%%%%%%%%%%%%%%%%%%%%%%%%%%%%%%%%%%
In principal, some of the non-trivial steady state solutions described 
before can be glued together to form discontinuous, periodic steady state
solutions~\cite{GKR03,Gre04}. Steady state solutions can be linked by a shock
wave, if the  quantities $\rho_{-}$, $v_{-}$ left to the interface and the
corresponding quantities $\rho_{+}$, $v_{+}$ right of the interface
satisfy the following conditions~\cite{AwR00,Gre01}:
\be
\label{shockcondition}
\rho_{+} > \rho_{-},
\ee
\be
\label{Temple}
\rho_{+} (v_{+} - u(\rho_{+})) = \rho_{-} (v_{-} - u(\rho_{-})).
\ee
With the equation of steady state solutions
\be
\rho_{\pm} v_{\pm} = q + \rho_{\pm} w,
\ee
we obtain from Eqn.~(\ref{Temple})
\be
\label{steadyshock}
w (\rho_{+} - \rho_{-}) = \rho_{+} u(\rho_{+}) - \rho_{-} u(\rho_{-}).
\ee
The monotonicity of the velocity of steady state solutions, which follows from
Eqn.~(\ref{ode}) (see also Fig.~\ref{fig3}), relates to the monotonicity
of the density according to
\be
\frac{d \rho}{dz} = - \frac{\rho}{v-w} \frac{d v}{dz}.
\ee
Therefore, condition~(\ref{shockcondition}) restricts periodic steady state
solutions linked by shock waves to regions II and IV of 
Fig.~\ref{fig2}. 

In the following we focus on steady state solutions of type CC in region II. In
particular, steady state solutions lying inside region II have to 
approach the jam line at velocity values for which the condition
\be
(\rho v^j(\rho))'' > 0
\ee
is fulfilled. For the chosen parameter values, this implies 
$\rho_{-} > 36.51$ [1/km/lane]. A necessary requirement to fulfill
Eqn.~(\ref{steadyshock}) is the condition 
\be
(\rho u(\rho))'|_{\rho=\rho_{-}} \ge (\rho v^j(\rho))'_{\rho=\rho_{-}},
\ee 
which results in our particular case to $\rho_{-} \le 61.57$ [1/km/lane]. 
Therefore, 
for the chosen parameter values there is only a very small parameter range
where periodic steady state solutions of type CC linked by shock waves are 
possible. However, in our numerical simulation, we find quasi steady state 
solutions in a much larger region of parameter space (see Fig.~\ref{fig1}).
\begin{figure}[htpb]
\includegraphics[width=\linewidth]{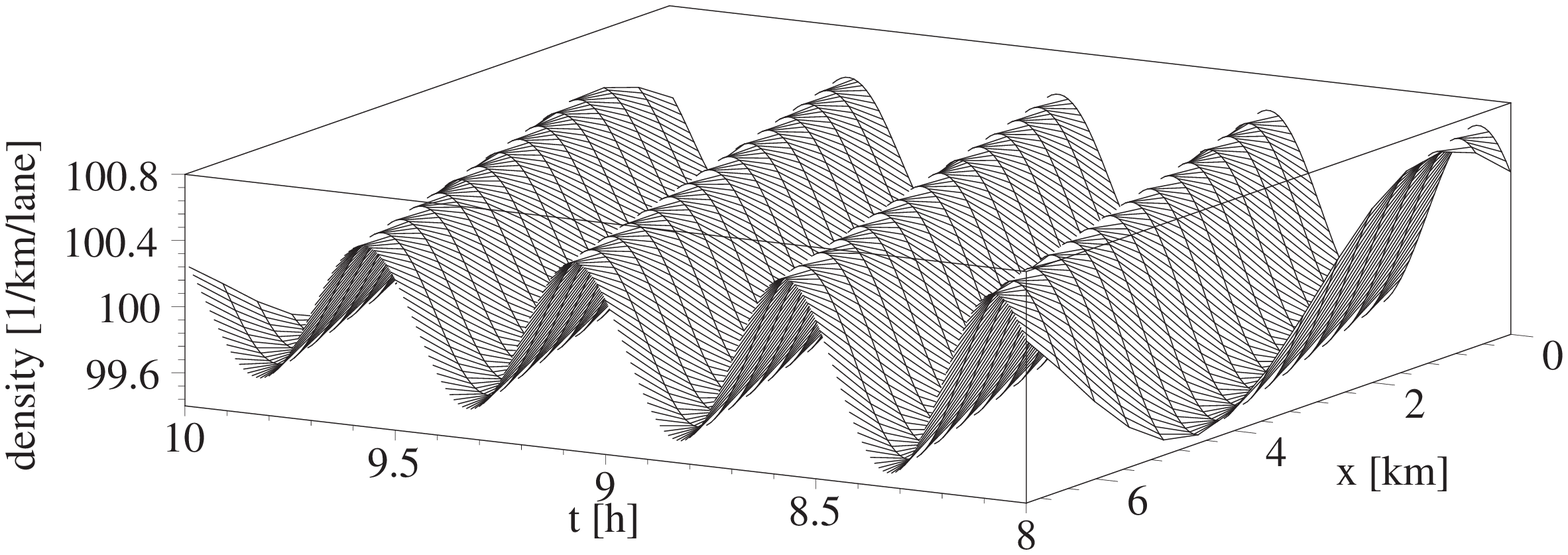}
\newline
\vspace{0.1cm}
\includegraphics[width=\linewidth]{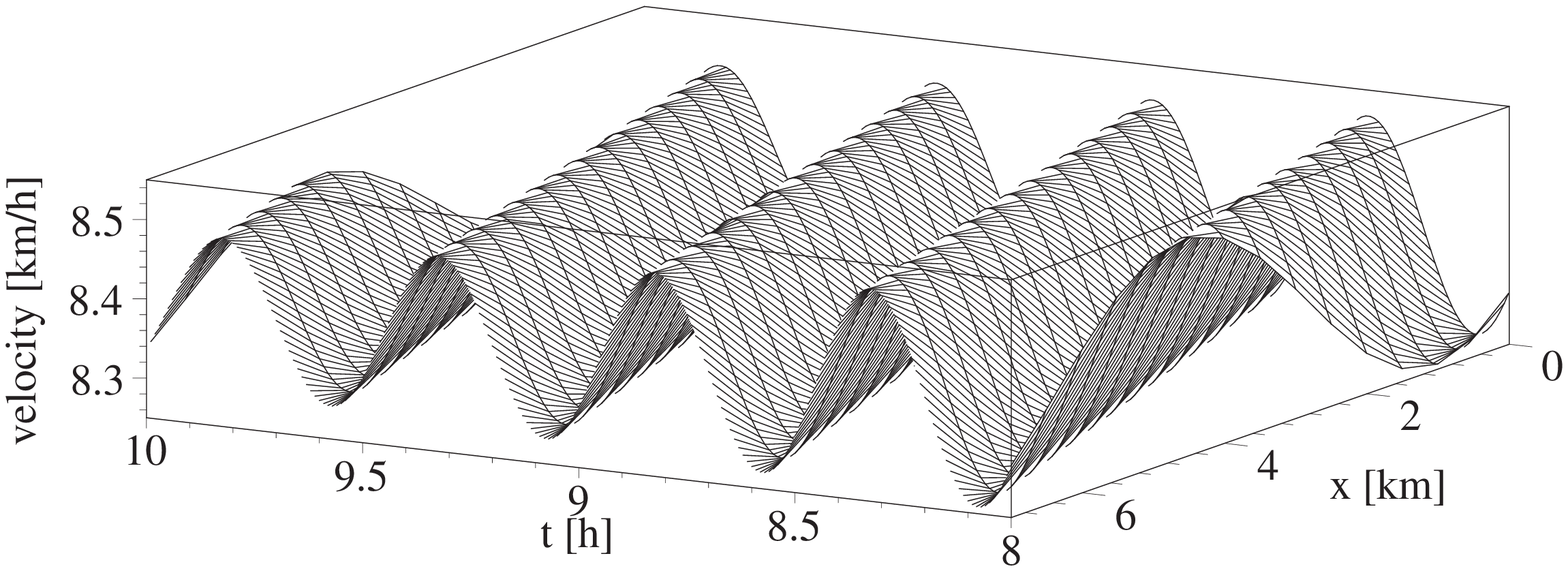}
\caption{Quasi steady state solution at the jam line for a numerical
  simulation  of perturbed equilibrium data with 
  initial density $\rho=100$ [1/km/lane]. Between a simulation time of 8 and 10 h 
  the evolution is quasi-stationary with a propagation speed of about 
  $w \approx -14$ km/h.
\label{fig5}}
\end{figure}
To obtain a more detailed picture of these quasi stationary solutions, we plot
the simulation results of unstable equilibrium data in Fig.~\ref{fig5}, 
focusing on the evolution between 8 and 10 hours simulation
time. We started the simulation of Fig.~\ref{fig5} with constant 
initial data $\rho=\rho_0=100$ [1/km/lane], $v = u(\rho_0)$ and a
sinusoidal density perturbation on top. After a long evolution time we 
obtain a quasi steady state close to the jam line, propagating with a velocity
$w \approx -14$ km/h upstream. This quasi steady state solution is not a true 
steady state solution, the amplitude slowly decreases with time, 
as it can be scarcely noted in the figure. The solution consists of
different branches close to the trivial steady state solutions lying in
regions I/II of Fig.~\ref{fig2}. Thus, for quasi steady state solutions the
quantities $q$ and $w$ of steady state solutions (see Eqn.~(\ref{stationary})) 
are only approximately constant. 

Similar quasi-steady state solutions also exist for a section of the high-flow
branch (see Sec.~\ref{sec:stability} below). They
consist of approximate steady state solutions lying in regions III/IV of 
Fig.~\ref{fig2}. These quasi steady state solutions travel
downstream with a velocity approximately given by the tangent of the high-flow
branch.
  
%%%%%%%%%%%%%%%%%%%%%%%%%%%%%%%%%%%%%%%%%%%%%%%%%%%%%%%%%%
\section{Stability analysis of the steady state solutions}
\label{sec:stability}
%%%%%%%%%%%%%%%%%%%%%%%%%%%%%%%%%%%%%%%%%%%%%%%%%%%%%%%%%%
Due to the importance of steady state solutions at the jam line and the
high-flow branch (see Fig.~\ref{fig1}) we focus the stability analysis
on the trivial steady state solutions of the BVT model. In principal the 
code can be applied to study the stability of the non-trivial solutions 
as well, although more accurate results may be obtained using specialized 
methods to this aim~(see e.g.~\cite{Boc04}).
%%%%%%%%%%%%%%%%%%%%%%%%%%%%%%%%%%%%%%
\subsection{Linear stability analysis}
%%%%%%%%%%%%%%%%%%%%%%%%%%%%%%%%%%%%%%
As presented in~\cite{SiM205} the equilibrium flow curve
$\rho v = \rho u(\rho)$ is linearly stable for $\rho<\rho_1$
and linearly unstable in the intermediate to high density regime 
$\rho_1<\rho<\rho_2$. 
Here we extend the linear stability analysis to all steady state solutions
obtained from setting $\beta(\rho,v)=0$, i.e. steady state solutions 
Eqs.~(\ref{jam}) and~(\ref{tip}). We denote the corresponding constant states 
$(\rho_0,v^{j/h}(\rho_0))$. 
Plugging the ansatz 
\bea
\rho & = & \rho_0 + \tilde{\rho} \exp(ilx+\omega(l)t), \\ 
v & = & v^{j/h}(\rho_0) + \tilde{v} \exp(ilx+\omega(l)t)
\eea
into the evolution equations~(\ref{rho}) and~(\ref{v}) we obtain as 
characteristic equations for the existence of solutions
$(\tilde{\rho},\tilde{v}) \ne (0,0)$:
\bea
\nonumber
(\omega + ilv^{j/h})^2  &+& (\omega + ilv^{j/h})
(ilu'\rho_0+(v^{j/h}-u)\frac{\partial \beta}{\partial v}) \\
& & - \rho_0 i l
\frac{\partial \beta}{\partial \rho}(v^{j/h}-u) = 0.
\eea
Solving the last equation for $\omega$ we can distinguish between
the linearly stable and unstable regime of the 
trivial steady-state solutions, i.e. curve sections with 
${\rm Re}(\omega) \le 0$ for arbitrary $l$ and curve sections for 
which ${\rm Re}(\omega) > 0$ for some $l$
 respectively. We find that the jam line is linearly unstable for 
$\rho_1 < \rho \le \tilde{\rho}_j = 39.73$ [1/km/lane] and linearly stable for
$\tilde{\rho}_j < \rho < \rho_2$, whereas the high-flow branch 
is linearly stable for densities $\rho_1 < \rho \le \tilde{\rho}_h =
39.73$ [1/km/lane] and linearly unstable for $\tilde{\rho}_h < \rho < \rho_2$.

In the following we will give a more intuitive explanation for the above
results. For trivial steady state solutions with exactly constant density 
and velocity profile, the characteristic
structure~(\ref{characteristiccondition}) does not give any 
restrictions. However, quasi steady state solutions with non-constant density 
(velocity) on these trivial steady state branches travel with a velocity 
$w$ which corresponds to the derivative of the flow density curve
\be
w \approx \frac{d (\rho v^{j/h})}{d\rho}.
\ee 
Hence, according to condition~(\ref{characteristiccondition}) the local
{\it characteristic cone} has to enclose the corresponding steady state branch
spanned by the characteristic speeds $\lambda_1$ and $\lambda_2$,
i.e. $\lambda_1 \le w \le \lambda_2$. For our parameter values
the characteristic condition restricts possible
stable (quasi) steady state solutions at the jam line to solutions fulfilling 
$\rho > \tilde{\rho}_j=39.73$ [1/km/lane] and those at the high-flow 
branch to solutions fulfilling $\rho \le \tilde{\rho}_h=39.73$
[1/km/lane] (see Fig.~\ref{fig2}). Note, that the two densities need not agree
for a general parameterization. For our parameterization, it follows 
from the appearance of the term $\Delta v(\rho)$ in both, Eqs.~(\ref{jam}) 
and (\ref{tip}). 
 
Next we study the nonlinear stability properties in simulations of the full
system.

\begin{figure}[htpb]
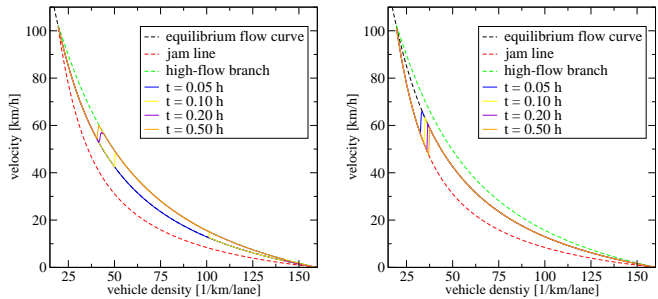

\vspace{0.4cm}
\begin{minipage}[t]{0.48\linewidth}
\includegraphics[width=1.0\linewidth]{fig6a.eps}
\end{minipage}
\hfill
\begin{minipage}[t]{0.48\linewidth}
\includegraphics[width=1.0\linewidth]{fig6b.eps}
\end{minipage}
\caption{(Color online) Left panel: 
Stability properties of the high-flow branch.
We study the stability properties by prescribing constant
steady state solutions $\rho = \rho_0$, $v = v^h(\rho_0)$ and on top a 
sinusoidal velocity perturbation  
$\delta v = v_{\rm{ampl}} \sin(\pi x)$ for $2<x<3$ km. In the figure, we plot 
the maximum value of $v^h(\rho_0) + v_{\rm ampl}$, $v_{\rm ampl} \le 0$, 
for which the initial data 
is unstable against the formation of synchronized flow, for different
evolution times. For densities $\rho < 40$ [1/km/lane], the high-flow branch 
is metastable, it becomes unstable against the formation of 
synchronized flow for sufficiently small values $v_{\rm ampl}$. 
For densities $\rho > 40$ [1/km/lane] the high-flow branch is unstable 
against the formation of synchronized traffic flow.\\
Right panel: Analysis of the stability properties of the jam line using 
constant
steady state solutions $\rho = \rho_0$, $v = v^j(\rho_0)$ and on top a 
sinusoidal velocity perturbation  
$\delta v = v_{\rm{ampl}} \sin(\pi x)$ for $2<x<3$ km. In the figure, 
we plot the minimum value of $v^j(\rho_0) + v_{\rm ampl}$, 
$v_{\rm ampl} \ge 0$, for which the 
initial data is unstable against the formation of synchronized flow, 
for different evolution times. For densities $\rho > 40$ [1/km/lane] the jam line
is metastable, it becomes unstable only for sufficiently large 
values $v_{\rm ampl}$. For densities $\rho < 40$ [1/km/lane] the jam line is 
unstable against the formation of synchronized traffic flow.  
\label{fig6}}
\end{figure}
%%%%%%%%%%%%%%%%%%%%%%%%%%%%%%
\subsection{Numerical results}
%%%%%%%%%%%%%%%%%%%%%%%%%%%%%%
We first analyze the stability properties of the high-flow branch
of steady state solutions, Eqn.~(\ref{tip}). To this aim, we use
constant steady state initial data $\rho = \rho_0$, $v = v^h(\rho_0)$ with a
sinusoidal perturbation $\delta v = v_{\rm{ampl}} \sin(\pi x)$ for $2<x<3$ km,
prescribing periodic boundary conditions on a 7 km long highway.
In order to decide whether synchronized flow appears during the numerical 
evolution, we use the criteria 
$v(\rho) < u(\rho)$ or $v(\rho) < v^h(\rho) - |v_{\rm ampl}|$.  

As our analysis shows, the stability properties of the high-flow branch 
depend on the particular perturbation. For the density regime $\rho_1 <
\rho \le \tilde{\rho}_{h}$, the high-flow branch is metastable against the
formation of synchronized flow, i.e. for small amplitude perturbations, no
synchronized flow appears, whereas for larger velocity perturbations with
negative amplitude $v_{\rm ampl}$, synchronized flow appears. For the density
regime $\tilde{\rho}_{h} < \rho < \rho_2$, the high-flow branch 
is unstable against the formation of synchronized flow. 
We summarize the corresponding results in the left panel of Fig.~\ref{fig6}. 

Second, we study the stability properties of the jam line, using constant 
steady state initial data $\rho = \rho_0$, $v = v^j(\rho_0)$ again with a
sinusoidal perturbation $\delta v = v_{\rm{ampl}} \sin(\pi x)$ for $2<x<3$ km.
In this case, we use the criteria $v(\rho) > u(\rho)$ or $v(\rho) > v^j(\rho)
+ |v_{\rm ampl}|$ to identify synchronized flow.

For densities $\tilde{\rho}_j < \rho < \rho_2$ the jam line is 
metastable against the formation of synchronized flow, i.e. for small 
amplitude perturbations, the jam line is stable, for larger amplitude 
perturbations (with positive amplitude $v_{\rm ampl})$, synchronized flow 
appears. In contrast, for densities $\rho_1 < \rho \le \tilde{\rho}_j$ [1/km] 
the jam line is unstable against the formation of synchronized flow. We show 
the results in the right panel of Fig.~\ref{fig6}.
 
\begin{figure}[htpb]
\vspace{1.0cm}
\includegraphics[width=0.9\linewidth]{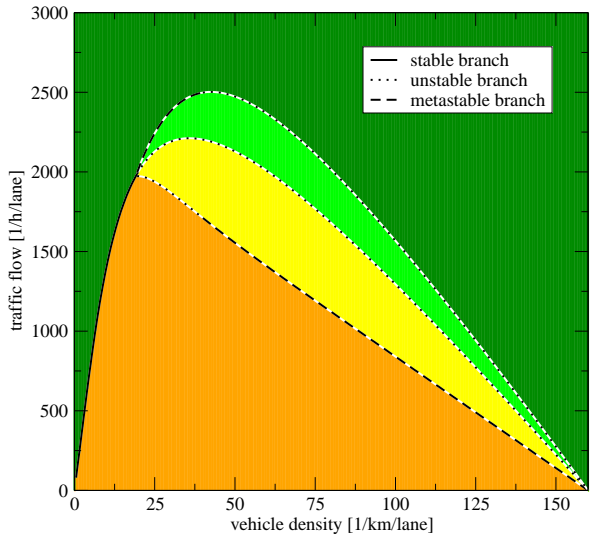}
\vspace{0.2cm}
\caption{(Color online) Results of the stability analysis of the trivial
  steady state solutions $\frac{d v}{d z} = 0$. Curve sections represented as 
  solid black line correspond to the linearly stable steady state solutions. 
  Curve sections represented as dotted line and dashed line correspond to 
  unstable and metastable steady state solutions respectively. 
\label{fig7}}
\end{figure}

We summarize the results of the stability analysis in 
Fig.~\ref{fig7}. The figure shows the observed gap between 
free and {\it congested flow} in the fundamental diagram of traffic flow 
separating the (meta-) stable branch sections.
 
%%%%%%%%%%%%%%%%%%%%%%%%%%%%%%%%%%%%%%%%%%%%%%%%%%%%%%%%%%%%%%%%%
\section{Identification of Kerner's three phases of traffic flow}
\label{sec:kerner}
%%%%%%%%%%%%%%%%%%%%%%%%%%%%%%%%%%%%%%%%%%%%%%%%%%%%%%%%%%%%%%%%%
Kerner~\cite{Ker04,KeR96,Ker98,Ker02} classifies traffic flow into 
three phases: free flow,
synchronized flow and wide moving jams. In this section we try to relate the
traffic states of the BVT model to Kerner's three phases. We will first
summarize our classification, before we will discuss the motivation.
\begin{itemize}
\item free flow:
Steady state solutions at the equilibrium velocity curve $u=u(\rho)$ for the 
density regime $0 \le \rho \le \rho_1$ (free equilibrium flow) and 
(quasi) steady state solutions close to the high-flow branch 
in the metastable regime $\rho_1 < \rho \le \tilde{\rho}_h$ make up 
the free flow state.
\item wide moving jams:
Spatially extended (quasi) steady state solutions at the jam line 
in the metastable regime  $\tilde{\rho}_j < \rho < \rho_2$ make up 
wide moving jams.
\item synchronized flow:
All other congested traffic states including the non-trivial steady state 
solutions of type BC and BD form synchronized traffic flow.
\end{itemize}
%%%%%%%%%%%%%%%%%%%%%%
\subsection{Free flow}
%%%%%%%%%%%%%%%%%%%%%%
For small densities ($0  \le \rho \le \rho_1$) free flow is stable in the BVT
model. Moreover, the model can reproduce the metastability of free flow 
against the formation of synchronized flow (see Fig.~\ref{fig6}), which is 
observed for traffic states at the high-flow branch 
$\rho_1 < \rho \le \tilde{\rho}_h$. In the model
instabilities only appear for velocity perturbations with negative amplitude. 
%%%%%%%%%%%%%%%%%%%%%%%%%%%%%
\subsection{Wide moving jams}
%%%%%%%%%%%%%%%%%%%%%%%%%%%%%
As our results show wide moving jams as defined above
are stable against small amplitude perturbations. For the chosen parameter
values the propagation speed of wide moving jams lies between $-16 < w < -14$
[km/h] for the density region $\tilde{\rho}_t < \rho \le \rho_2$, 
i.e. it is nearly constant and reproduces the observed value.

\begin{figure}[h!]
\noindent
\begin{minipage}[t]{\linewidth}
\includegraphics[width=\linewidth]{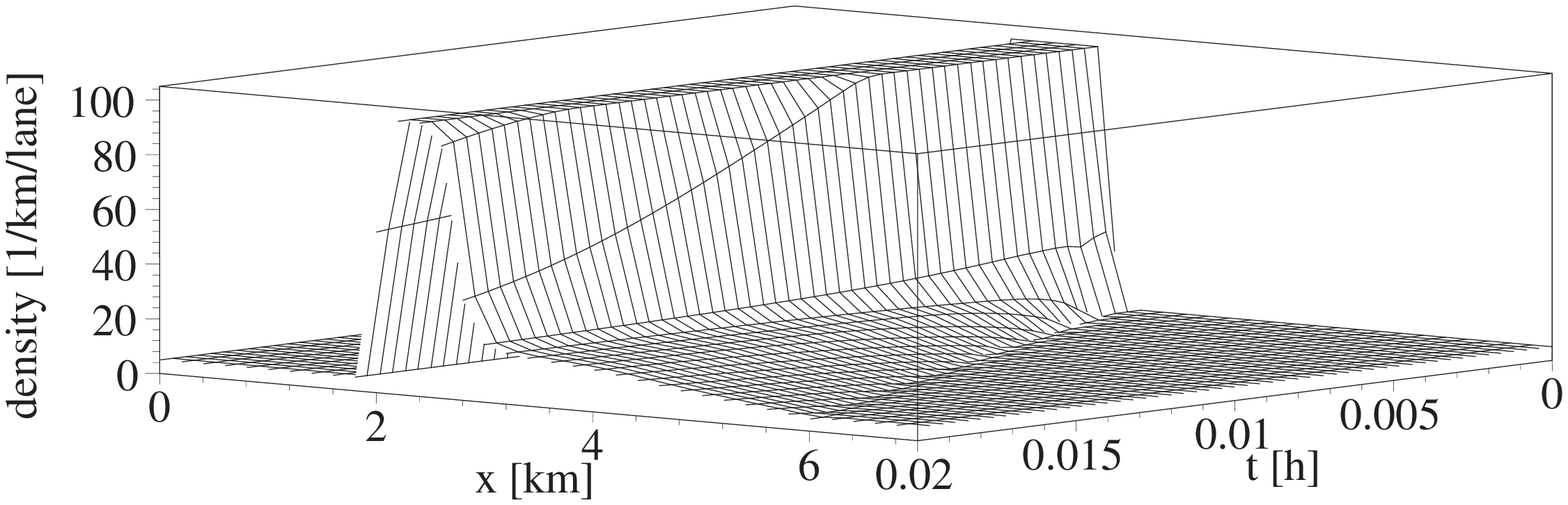}
\vspace{0.2cm}\\
\includegraphics[width=\linewidth]{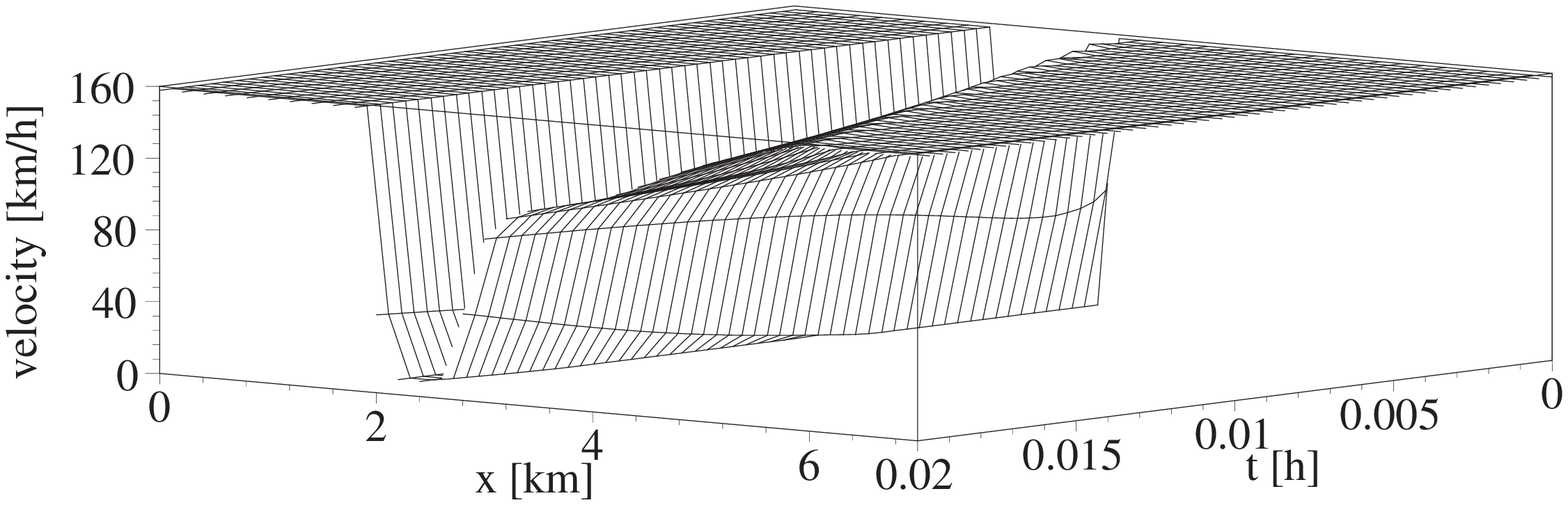}
\end{minipage}
\vspace{0.7cm}\\
\begin{minipage}[t]{.48\linewidth}  
\centering\epsfig{figure=fig8c.eps,width=\linewidth}
\end{minipage}\hfill
\begin{minipage}[t]{.48\linewidth}
\centering\epsfig{figure=fig8d.eps,width=\linewidth}
\end{minipage}
\caption{(Color online)
Upper two panels: Simulation of the outflow from a wide moving jam. Plotted
  are the evolution of the density and the velocity as a function of space and
  time. In the initial data we prescribe a wide moving jam with density
  $\rho_j=100$ [1/km/lane] between $2$ and $3$ km, and free flow with density
  $\rho_f=5$ [1/km/lane] elsewhere. During the evolution, the wide moving jam 
  narrows down and finally dissolves.\\
Lower left panel: Flow-density diagram for the above simulation results 
after $t=0.02$ h. In addition we plot the curve
representing the equilibrium flow as well as the jam line and the high-flow
branch. We determine the outflow of the wide moving jam at
that point for which the velocity of the outflow differs from the
  equilibrium velocity by less than 1 \%.\\
Lower right panel:
Outflow from wide moving jams, varying the jam density $\rho_j$ of the wide
  moving jams between $2$ and $3$ km. The three different curves correspond to
different values of free flow density $\rho_f$ in the region between $0$ and
$2$ km and $3$ and $7$ km. As one can see from the plot, the outflow from 
the wide moving jams only varies within a very small range of flow values  
and it is largely independent of the density of free flow. The typical outflow 
from wide moving jams for the chosen parameter values is 
$f_{\rm{out}} \approx 1914$ [1/h/lane], which is far below the maximum of 
metastable free flow $f = 2487$ [1/h/lane].}
\label{fig8}
\end{figure}

We further analyzed the outflow from wide moving jams (see Fig.~\ref{fig8}). 
We find for wide moving jams in the density region 
$\tilde{\rho}_t < \rho < 125$ [1/km/lane] a constant outflow of about 
$f_{\rm{out}} = 1914$ [1/h/lane]. 
This can be seen in the lower right panel of 
Fig.~\ref{fig8}, where we plot the outflow from wide moving jams
$\rho = \rho_j, v  =  v^j(\rho_j)$ initially located between 2 and 3 km 
and surrounded by a  region of free flow $\rho = \rho_f, v  =  u(\rho_f)$.
%%%%%%%%%%%%%%%%%%%%%%%%%%%%%% 
\subsection{Synchronized flow}
\begin{figure}[htpb]
\vspace{1.0cm}
\includegraphics[width=0.9\linewidth]{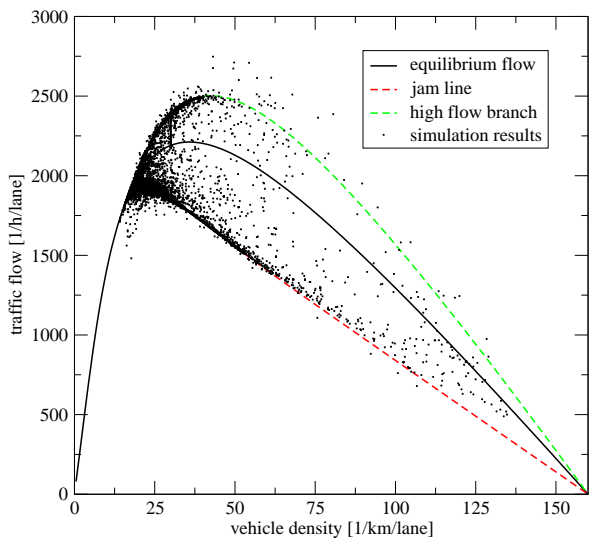}
\vspace{0.5cm}
\caption{(Color online) Formation of synchronized flow from metastable free flow of density
  $\rho=30$ [1/km/lane] with an initial velocity perturbation $\delta v = - 7
  \sin(\pi x)$ located between 2 an 3 km. Due to the velocity perturbation,
the free flow state breaks down, leading to a complicated pattern of 
synchronized flow which covers a wide region of states in the fundamental
  diagram. Finally, moving jams form, which can in turn lead to free flow
  of lower density, thus reproducing the hysteresis effect observed in 
  traffic dynamics. The plot shows all data points corresponding to the
  constant time slices at $t = i \Delta t$, where $\Delta t = 0.1$ h, 
$i = 0, .., 50$. 
\label{fig9}}
\vspace{0.4cm}
\end{figure}

%%%%%%%%%%%%%%%%%%%%%%%%%%%%%%
Synchronized flow as defined above covers a wide region in the fundamental
diagram. This can be already seen in the left panel of Fig.~\ref{fig4} for
steady state solutions of type BC and BD. We exemplarily show the formation 
of synchronized flow from free flow of density $\rho = 30$ [1/km/lane]. As a 
nucleus for the emergence of synchronized flow, we use a velocity perturbation
$\delta v = - 7 \sin(\pi x)$ located between 2 an 3 km on a highway with
periodic boundary conditions. The evolution of these initial data leads to 
states which are widely scattered in the fundamental diagram, as it can 
be seen in Fig.~\ref{fig9}. We remark that in an earlier work~\cite{TrH99} the 
scattering was reproduced by modeling different vehicle types. Here, in
contrast, the scattering already follows from the traffic dynamics without 
distinguishing between different vehicle types. Including different vehicle
types into the BVT model, which is beyond the scope of the current work, would
be expected to further widen the scattering.
%%%%%%%%%%%%%%%%%%%%%%%%%%%%%%%%%%%%
\section{Traffic flow at bottlenecks}
\label{sec:bottleneck}
\begin{figure}[htpb]
\noindent
\epsfig{figure=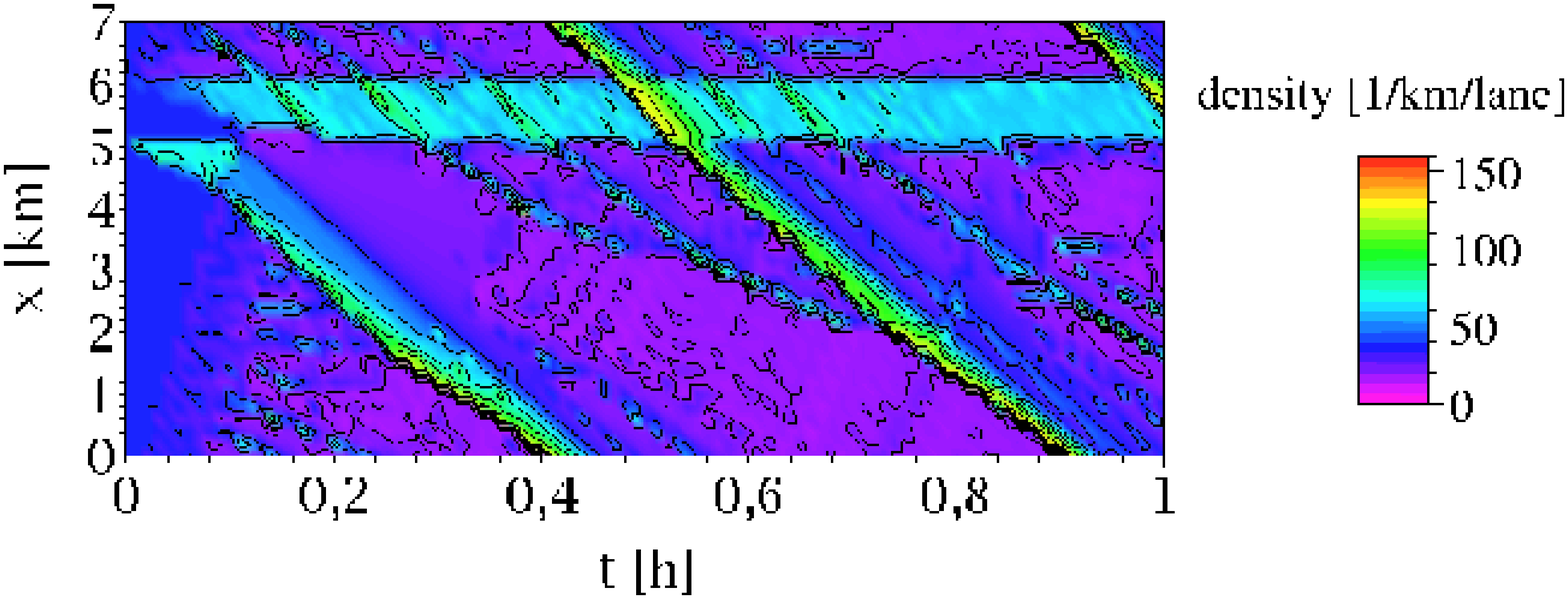,width=\linewidth}
\epsfig{figure=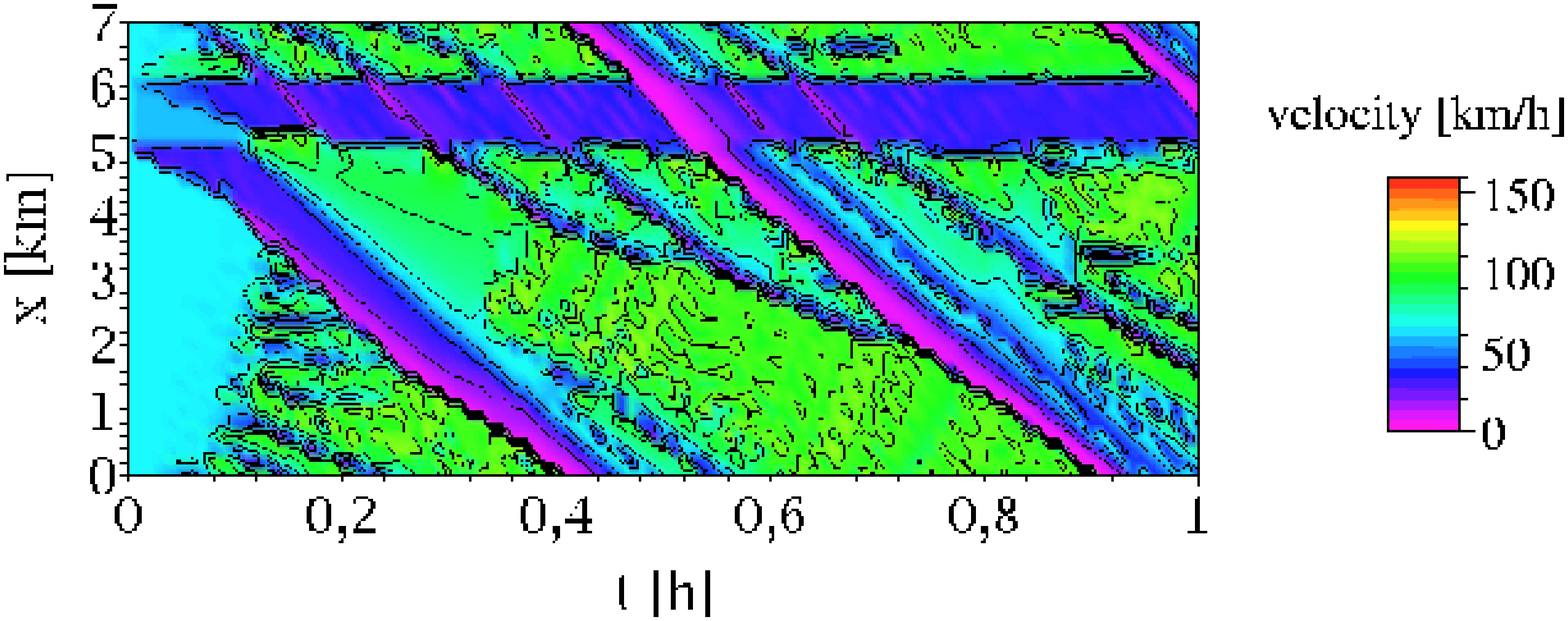,width=\linewidth}
\caption{(Color online) Formation and propagation of wide moving jams. The upper panel shows
  the evolution of the density, whereas the lower panel shows the evolution of
  the velocity. At the bottleneck located between 5 and 6 km, 
  synchronized flow forms, which finally leads to a wide moving
  jam. This wide moving jam moves with a velocity of about -15
  km/h (i.e. upstream) and swallows moving jams during this propagation. It
  further travels through the bottleneck. See the text for a detailed description.
\label{fig10}}
\end{figure}

%%%%%%%%%%%%%%%%%%%%%%%%%%%%%%%%%%%%
In this section we study the behavior of traffic flow in the BVT model at a
highway bottleneck. We focus the discussion 
on two simulation runs of a two-lane highway with periodic boundary 
conditions~\footnote{Using
  periodic boundary conditions enables us to study the propagation of moving
  jams through a bottleneck modeling only a single bottleneck.}. Again, we use
a longitudinal extension of the highway of 7 km with homogeneous 
initial free flow of density $\rho = 37.5$ [1/km/lane]. We model the bottleneck
simply by a velocity modification (velocity drop for free flow) 
between 5 and 6 km, setting the velocity to a modified value $v^{\rm{mod}}$ 
after each numerical evolution step,
\be
\label{bottleneck}
v^{\rm{mod}} = v + (u(\rho) - v -0.1 {\rm \ km/h}) |\sin(\pi x)|
\ee 
We show the numerical evolution in Fig.~\ref{fig10}.
Despite the simplicity of the initial setup the numerical evolution shows a
very complicated dynamics. As we will discuss below, we observe the formation
of synchronized flow and wide moving jams.

As a consequence of the bottleneck, the initial velocity drops to smaller 
values in the bottleneck region (dark blue regions between 5 and 6 km in 
the velocity plot), but also further upstream (dark blue regions between
4 and 5 km at about 0.1h). Both regions correspond to synchronized flow.
The first synchronized flow region stays fixed at the
bottleneck, however the upstream front can oscillate in time (e.g. between 
0.8 h and 0.9 h). The second region of synchronized flow travels
upstream. It takes some time until an accentuated wide moving jam 
with velocities close to zero forms. This wide moving jam travels further
upstream and reenters the numerical domain at 7 km after $t \approx 0.4$ h
due to the periodic boundary conditions used in the numerical simulation.
When reaching the bottleneck, it simply travels through the first synchronized
region, thus becoming a foreign wide moving jam. Note, that the velocity of the
downstream front of this wide moving jam is nearly constant and has a value of
about -15 km/h.

Between the wide moving jams we observe regions of low density and high
velocity, which correspond to free flow (see e.g. the region at $x = 1$ km
for $t = 0.7$ h), and smaller moving jams. As one can see from the plot there 
are several regions (pinch regions) where these additional moving jams form 
(see e.g. the region between 0 and 3 km for $t \approx 0.15$ h or at about 
1.5 km for $t \approx 0.45$ h). 
For these moving jams, the downstream front is in general not as robust
as for the wide moving jam described above. At $x = 3.5$ km for $t = 0.55$ h,
we observe the merging of two moving jams, which are finally swallowed by the
wide moving jam at about $x = 2$ km for $t = 0.75$ h. We also observe an
example for the catch effect of a narrow moving jam, see the region at $x =
5.5$ km for $t \approx 0.15$ h.

\begin{figure}[htpb]
\noindent
\epsfig{figure=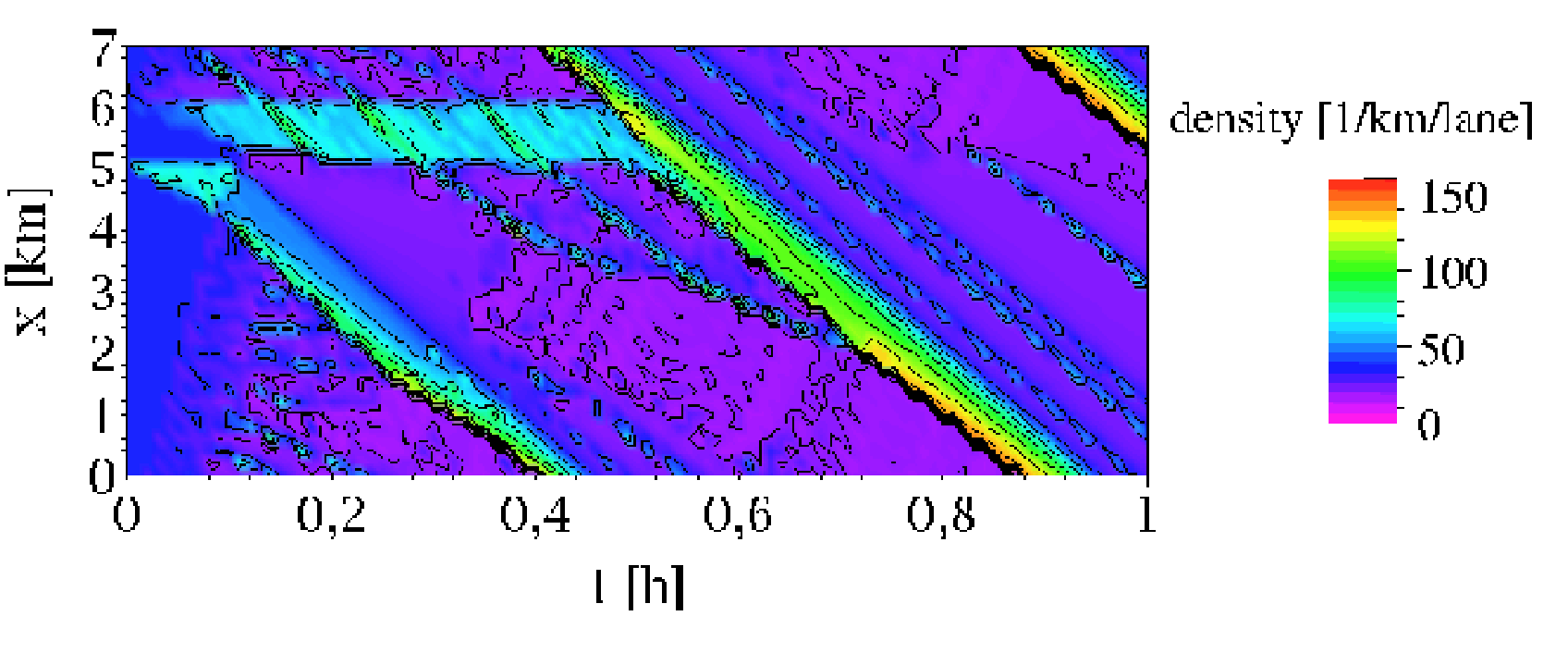,width=\linewidth}
\epsfig{figure=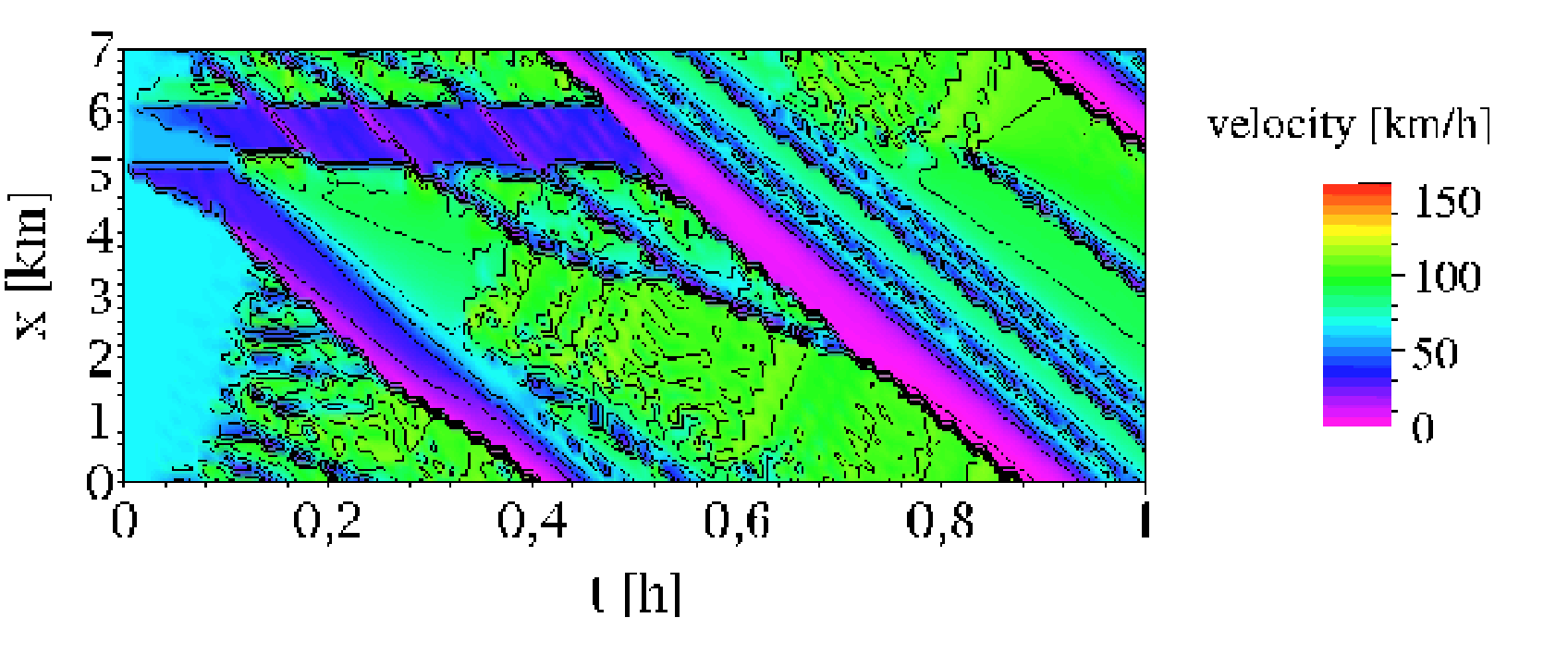,width=\linewidth}
\caption{(Color online) Formation and propagation of wide moving jams. The simulation setup
  is identical to that of Fig.~\ref{fig10}, except that after a time of 0.5 h
  the bottleneck is not effective any longer. As a consequence, the 
  synchronized flow region  pinned to the bottleneck region disappears, the 
  corresponding wide moving jam becomes wider at late times.\label{fig11}}
\end{figure}
\begin{figure}[htpb]
\vspace{1cm}
\noindent
\epsfig{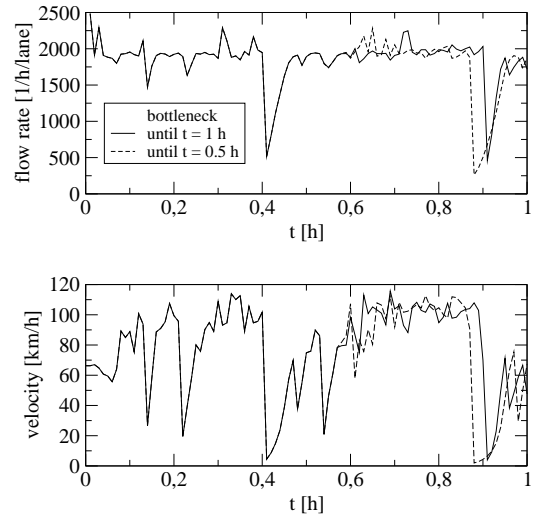}
\vspace{1cm}
\caption{Time series of the flow rate (upper panel) and the velocity (lower
  panel) as measured by a detector at location $x = 0$ km for the simulation
  results of Figs.~\ref{fig10} and~\ref{fig11}. Inside the wide moving jam,
  small values of the traffic flow and in particular the velocity are reached.
\label{fig12}}
\end{figure}

Figure~\ref{fig11} shows the simulation results for the same initial setup, 
except that the bottleneck Eqn.~(\ref{bottleneck}) is only effective 
for simulation times $t < 0.5$ h. As a consequence the synchronized flow region
pinned to the bottleneck disappears at later times. The wide moving jam 
traveling through
the former bottleneck region expands and reaches velocities close to zero
inside the jam. This can be also seen in Fig.~\ref{fig12} where we plot the 
time series of the flow rate and the velocity for a detector located at $x =
0$ km. 

%%%%%%%%%%%%%%%%%%%%%%
\section{Discussion}
\label{sec:discussion}
%%%%%%%%%%%%%%%%%%%%%%
The BVT model is a macroscopic, deterministic model, which 
describes vehicular traffic flow using standard methods from continuum fluid
dynamics. It uses an equilibrium flow-density curve. In contrast to earlier
models, however, the parameter range of the effective relaxation coefficient is
extended to negative values. As a consequence, the equilibrium flow curve does
not describe traffic states in the congested regime directly, but still 
determines the characteristic structure of the model. 
An additional consequence of the
negative effective relaxation coefficient is the appearance of (two) additional
branches of trivial steady states. The characteristic structure, i.e. the
finiteness of propagation speeds, restricts the
stability of these steady state solutions. The high-flow branch is metastable 
against the formation of synchronized flow for intermediate densities and 
unstable for high densities. We interpret the metastable section of high-flow
branch as metastable free flow. Stable free flow, in contrast, corresponds
to the stable equilibrium flow. The low flow branch in the congested regime
(i.e. the jam line) is
unstable against the formation of synchronized flow for intermediate densities
and metastable for high densities. We interpret spatially extended solutions
at the metastable branch of the jam line as wide moving jams. We further
identify the unstable sections of the high-flow branch and the jam
line, as well as the additional (steady state) solutions in the congested
regime, which can lead to very complicated oscillatory patterns, as 
synchronized flow. Thus, synchronized flow covers a wide region of congested 
states in the fundamental diagram, without distinguishing between different 
vehicle types and driver characteristics in the model.

There are some additional results supporting the BVT model. In
particular, the model ensures that wide moving jams do not form spontaneously
from free flow. When the velocity drops below the critical value in free flow,
the velocity is driven to even smaller values at that location which results
in a strong gradient in the velocity. It is only after complicated
oscillations have occurred, which lead to a rearrangement of the density and 
the velocity, that an extended steady state solution close to the jam line, 
i.e. a wide moving jam, appears. 

We further can reproduce the characteristic properties of real wide moving jams
with our model. For the chosen parameter values, the downstream
front of wide moving jams travels upstream with a nearly constant velocity 
of about 15 km/h. Moreover, the outflow from wide moving jams is largely 
independent of the characteristics of the wide moving jam. We obtain a 
typical outflow of 1914 vehicles/h/lane. Furthermore, we showed, that wide 
moving jams travel through bottlenecks, whereas smaller moving jams can be 
caught by a bottleneck.  

%%%%%%%%%%%%%%%%%%%%%%%%%%%
\section*{Acknowledgments}
%%%%%%%%%%%%%%%%%%%%%%%%%%%
We thank James Greenberg for encouraging comments.

\end{document}